\documentclass[10]{nature}
\usepackage{subcaption}
\usepackage{float}
\usepackage{graphicx} 
\usepackage{amsmath,xcolor,colortbl,multirow}
\usepackage{amssymb,soul,color}
\usepackage[left]{lineno}
\usepackage[version=3]{mhchem}
\usepackage{comment}
\usepackage{hhline}
\usepackage{fancyhdr}
\DeclareUnicodeCharacter{0301}{}
\usepackage{parskip} 
\usepackage{textcomp}
\usepackage[utf8]{inputenc}

\providecommand{\keywords}[1]{\textbf{Keywords: } #1} 

\begin{document}
\title{Computational Screening and Discovery of Silver–Indium \\ Halide Double Salts.}
\author{Christos Tyrpenou~$^{\dagger}$, G. Krishnamurthy Grandhi~$^{\ddagger}$,  Paola Vivo~$^{\ddagger}$, Mikaël Kepenekian~$^{\dagger,*}$, George Volonakis~$^{\dagger,*}$}

\maketitle

\vspace{0.2cm}
\begin{affiliations}
\item ${\dagger}$ Univ Rennes, ENSCR, CNRS, ISCR (Institut des Sciences   Chimiques de Rennes), UMR 6226, France.
\item ${\ddagger}$ Hybrid Solar Cells, Faculty of Engineering and Natural Sciences, Tampere University, P.O. Box 541, FI-33014, Tampere, Finland.

Email Address: yorgos.volonakis@univ-rennes.fr

\end{affiliations}

\begin{abstract}
Perovskite-inspired materials have emerged as promising candidates for both outdoor and indoor photovoltaic applications owing to their favorable optoelectronic properties and reduced toxicity. Here, we employ the experimentally realized AgBiI$_4$ double salt as a structural prototype and replace Bi$^{3+}$ with In$^{3+}$ to design a novel lead-free halide compound, AgInI$_4$. First-principles calculations predict that AgInI$_4$ is both chemically and dynamically stable, exhibiting a direct band gap of 1.72 eV, comparable to its bismuth analogue. However, its predicted photovoltaic performance, evaluated using the spectroscopic limited maximum efficiency metric, is lower under both solar and LED illumination. This reduction arises primarily from symmetry-forbidden optical transitions and the absence of Bi-derived 6s$^2$ lone-pair states at the valence band maximum. High-throughput screening of the Ag–In–I ternary phase-space reveals several more stable and metastable compounds that fall into two structural families: tetrahedrally and octahedrally coordinated, with characteristic band gaps near 3.0 eV and 2.0 eV, respectively. Despite multiple synthetic attempts, the predicted AgInI$_4$ phase could not be experimentally realized, underscoring the challenges of stabilizing indium-based halide double salts. While these materials are unlikely to serve as efficient photovoltaic absorbers, their tunable band gaps and stability make them promising candidates for charge transport and other optoelectronic applications.
\end{abstract}

\vspace{2cm}


\clearpage
\section{Introduction}

Over the past few years, substantial efforts have been made to identify lead-free perovskites that would retain the remarkable optoelectronic properties of record-breaking lead-based perovskites \cite{WeiPeng2023, XiaodongLi2022, ZhenLi2022, JiangQi2022, LiangZheng2023} while enhancing the materials stability and reduce their environmental impact. These exceptional properties exhibited by Pb-based perovskites have been attributed to the unique electronic configuration of Pb$^{2+}$, and in particular the role of the 6s$^2$ lone pair \cite{fabini2020, Labrecht2013}. Consequently, the most obvious strategy explored was to replace Pb with atoms belonging to the same group in the periodic table, like divalent tin (Sn$^{2+}$) and germanium (Ge$^{2+}$). Yet, Sn and Ge based perovskites have poor stability due to the susceptibility of Sn$^{2+}$ and Ge$^{2+}$ to oxidize in the +4 state instead of the desired +2 state, especially under ambient conditions \cite{FengHao2014, Noel2014}. Furthermore, in Sn-, and even more in Ge-based perovskites, the stereochemical activity of the s$^2$ lone pair results to  structural distortions of the octahedra in the lattice, hence impacts the electronic structure and the materials optical properties \cite{Fabini2016,Malavasi2021,LiuYang2022,Balvanz2024}. It is also worth noting that the presence of Sn can also raise environmental concerns, further complicating the search for totally benign alternatives \cite{Babayigit2016}. An alternative strategy for designing environmentally friendly perovskites involves replacing two lead (Pb) atoms in conventional ABX$_3$ lattice with two elements in distinct oxidation states: a monovalent element (M$^{1+}$) and a trivalent element (M$^{\prime}$$^{3+}$). This substitution leads to the formation of the so-called double perovskite lattice with the stoichiometric formula A$_2$MM$^\prime$X$_6$, expanding the range of potential candidate materials \cite{bera2022, yang2018}. A notable compound within this class of materials was discovered in 2016 is Cs$_2$AgBiBr$_6$ \cite{Volonakis2016,Slavney2016,Lei2021} that exhibits an indirect band gap in the visible, and has since attracted considerable attention for their optoelectronic applications, especially in emerging solar cells \cite{WangBaoning2021,XiaoqingYang2020}. These work lead to the computational discovery of Cs$_2$AgInCl$_6$ in 2017, another double perovskite~\cite{Volonakis2017}, which has also attracted considerable interest from the research community due to its direct bandgap and has also been as a promising candidate for optoelectronic applications like white-light emitters \cite{HaiyanWang2024,Alivisatos2019,LuoJiajun2018}.

Following halide double perovskites, another class of materials has emerged as promising alternatives, the so-called halide double salts (also called rudorffites), which are part of the broader family of perovskite inspired materials (PIMs). A notable example is the Ag/Bi halide double salt, represented by the stoichiometric formula Ag$_a$Bi$_b$I$_x$ (where x = a + 3b), which has been identified as a highly stable compound class with significant potential for solar cell applications. These materials are also easy to process and have been successfully synthesized by various methods \cite{Kim2016, Khazaee2019, Danilovic2020, Ye2020, Prasad2021}. They are essentially mixtures of the two salts, for example AgI and BiI$_3$ in the case of AgBiI$_4$, and exhibit well-positioned optical band gaps in the visible region, ranging from 1.4 to 2.0 eV \cite{LanWang2021}. In particular, AgBiI$_4$ and Ag$_3$BiI$_6$ have shown exceptional potential for photovoltaic device fabrication, and solar cell devices have attained high short circuit current of above 10 mA cm$^{-2}$ \cite{Turkevych2021,Turkevych2017} and photovoltaic efficiencies of up to 5.6\% \cite{Pai2019}. Considering the rapid growth of Internet of Things (IoT) devices, their requirement for autonomous and sustainable operation along with the necessity to employ light absorbing materials  with band gaps precisely tailored to the indoor illumination spectrum (Figure S3) ($\approx$ 1.8 – 2.0 eV) \cite{Grandhi2025,Grandhi2023,Burwell2024,Panda2025} PIMs exhibit bandgaps close to the optimal range for indoor operating conditions, with Ag$_3$BiI$_6$ reaching an indoors efficiency of 5.17\% \cite{Turkevych2021}, while other closely related PIMs like Cs$_2$AgBi$_2$I$_9$ and CsMAFA-Sb:Bi, achieved 7.6\% \cite{Krishnaiah2025} and 10.11\% \cite{Lamminen2025} under LED illumination conditions, respectively. 

However, despite their most promising properties as light absorbers, these materials are particularly challenging to model as they can contain partially occupied atomic positions. Here, we employ a reduced-symmetry model using Wyckoff position splitting we have previously developed~\cite{Cucco2022}, to explore halide double salts with Ag and In, to design a direct band-gap material, analogous to Cs$_2$AgInCl$_6$. Substituting Bi$^{3+}$ with In$^{3+}$ in the reduced-symmetry model of AgBiI$_4$, our first-principles calculations predict that AgInI$_4$ has a direct bandgap within the visible range, suggesting a potential advantage for the In-based material for thin-film photovoltaic applications compared to the Ag-Bi iodides, which exhibit indirect bandgaps. We further employ the spectroscopic limited maximum efficiency (SLME) to assess the outdoors and indoors photovoltaic performance of the In-based double salt in comparision with AgBiI$_4$. Finally, we move on to explore the complete Ag-In-I phase space to indentigy the potential existence of other stable polymorphs, by developing a systematic materials screening process within the Materials Project database \cite{Jain2013}. We successfully identify phases that are more stable than the reduce-symmetry model AgInI$_4$. These newly identified polymorphs belong into two sets of materials with different coordinations for Ag and In, which exhibit distinct bandgaps, highlighting the versatility of this phase space for optoelectronic applications.

\section{Results and Discussion}
\subsection{Structural and Electronic Properties}\

\begin{figure}[H]
    \centering
    \includegraphics[width=1.0\textwidth]{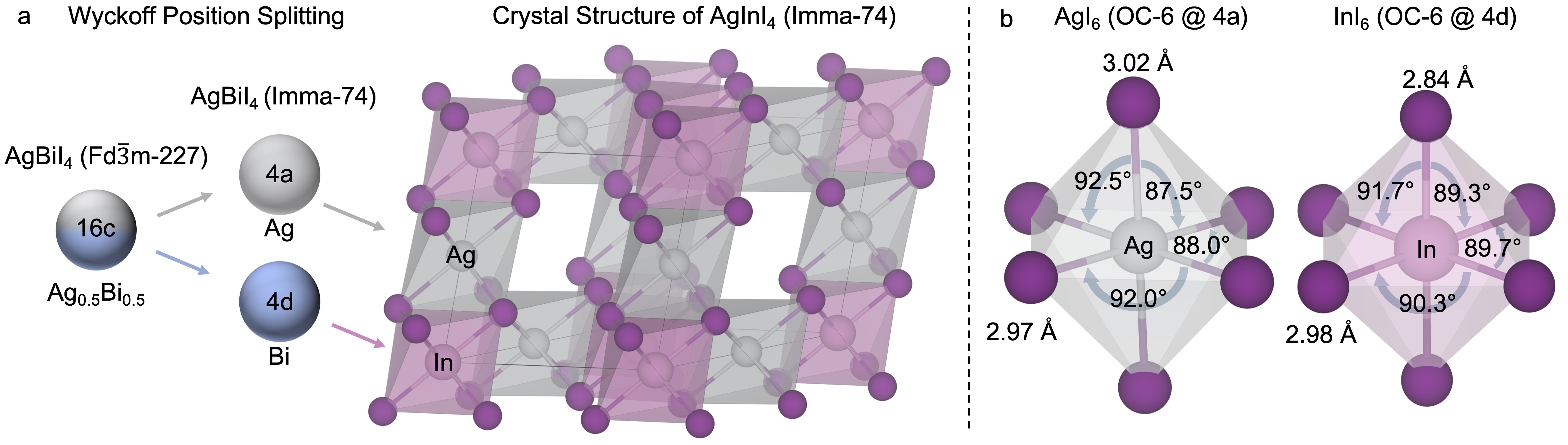}
    \caption{a) Crystal structure of AgInI$_4$ based on the reduced symmetry model of AgBiI$_4$, b) alongside the distorted AgI$_6$ and InI$_6$ octahedra (OC-6) of the AgInI$_4$ structure. The OC-6 abbreviation refers to the IUPAC symbol for octahedron coordination environment.}
    \label{fig:Structure}
\end{figure}

The double perovskite Cs$_2$AgBiCl$_6$ exhibits an indirect band gap originating from the strong directional interactions between Ag-\emph{d} and Bi-\emph{s} orbitals at the top of the valence band~\cite{Volonakis2016}. Substituting Bi$^{3+}$ ([Xe] 4f$^{14}$5d$^{10}$6s$^2$) with In$^{3+}$ ([Kr]4d$^{10}$) to form Cs$_2$AgInCl$_6$ results in a direct band gap material, as the valence band maximum no longer contains \emph{s}-type orbital contributions~\cite{Volonakis2017}. Following this materials design principle, we adopt a reduced-symmetry model of the AgBiI$_4$ ternary compound obtained using the Wyckoff position splitting method~\cite{Cucco2022}. This approach, which exploits group–subgroup relations derived from the parent space group, enables the systematic reduction of crystal symmetry and the corresponding assignment of the partially occupied atomic positions in a controlled manner (see Figure \ref{fig:Structure}a). In this way, the structure can be described using an alternative unit cell that retains most of the symmetry operations of the original lattice. We employ this cell to replace the B-site bismuth (Bi$^{3+}$) atom with indium (In$^{3+}$) and examine the resulting structural and electronic modifications induced by this substitution.

The crystal lattice of AgInI$_4$ comprises distorted AgI$_6$ and InI$_6$ octahedra, shown in Figure \ref{fig:Structure}b. Within the \textit{Imma} subgroup (space group 74), the A-site Ag atom occupies the central 4a position, whereas the B-site In atom resides at the corner 4d position of the reduced-symmetry model, forming a distinct edge-sharing coordination environment for Ag and In, as illustrated in Figure~\ref{fig:Structure}a. Similar to AgBiI$_4$, the lattice also contains vacant A- and B-sites. The structure was fully optimized (i.e., atomic positions and lattice parameters) using the DFT-PBEsol functional, yielding a lattice constant of $a = 8.418$~\AA\ for the Ag/In ternary compound. Analysis of the reduced-symmetry structure reveals notable differences in bond lengths at both the A- and B-sites. The average Ag–I bond length at the A-site is 2.98~\AA, slightly shorter than the experimentally reported average Ag–I bond length of 3.06~\AA\ for AgBiI$_4$ by Sansom \textit{et al.}~\cite{Sansom2017}. At the B-site, the In–I bond lengths are 2.98~\AA\ in-plane and 2.84~\AA\ out-of-plane (Figure~\ref{fig:Structure}b), both shorter than the corresponding Bi–I bond lengths (3.09–3.14~\AA) reported for AgBiI$_4$~\cite{Sansom2017,Barone2022}. These shorter bonds can be attributed to the considerably smaller ionic radius of In$^{3+}$ (0.80~\AA) compared to Bi$^{3+}$ (1.03~\AA). We evaluated the thermodynamic stability of AgInI$_4$ by calculating its formation enthalpy, defined as:
\begin{equation}
\Delta H = E_{\mathrm{AMX_4}} - (E_{\mathrm{AX}} + E_{\mathrm{MX_3}}),
\end{equation}
where $E_{\mathrm{AMX_4}}$, $E_{\mathrm{AX}}$, and $E_{\mathrm{MX_3}}$ are the total energies of AgInI$_4$, AgI, and InI$_3$, respectively. The calculated formation enthalpy of $-12$~meV/atom indicates that AgInI$_4$ is thermodynamically stable against decomposition into its binary precursors. This finding contrasts with a previous theoretical report \cite{HuanhuanLi2024}, which predicted AgInI$_4$ to be unstable. The discrepancy underscores the sensitivity of the computed stability to the precise atomic arrangement, particularly in systems derived from partially occupied atomic positions like the halide double salts. To further assess the material’s dynamical stability, we computed its phonon dispersion spectrum (Figure~S3). The absence of imaginary frequencies confirms that the structure is dynamically stable. Phonon calculations (Figure S3) show no imaginary frequencies, confirming the structure's dynamical stability. Coupled with the small driving force ($\Delta$H $\approx$ -12 meV/atom), this suggests a narrow kinetic window for experimental crystallization, consistent with our preliminary synthesis outcomes discussed in the following sections and Supplementary Information.

\begin{figure}[H]
    \centering
    \includegraphics[width=0.70\linewidth]{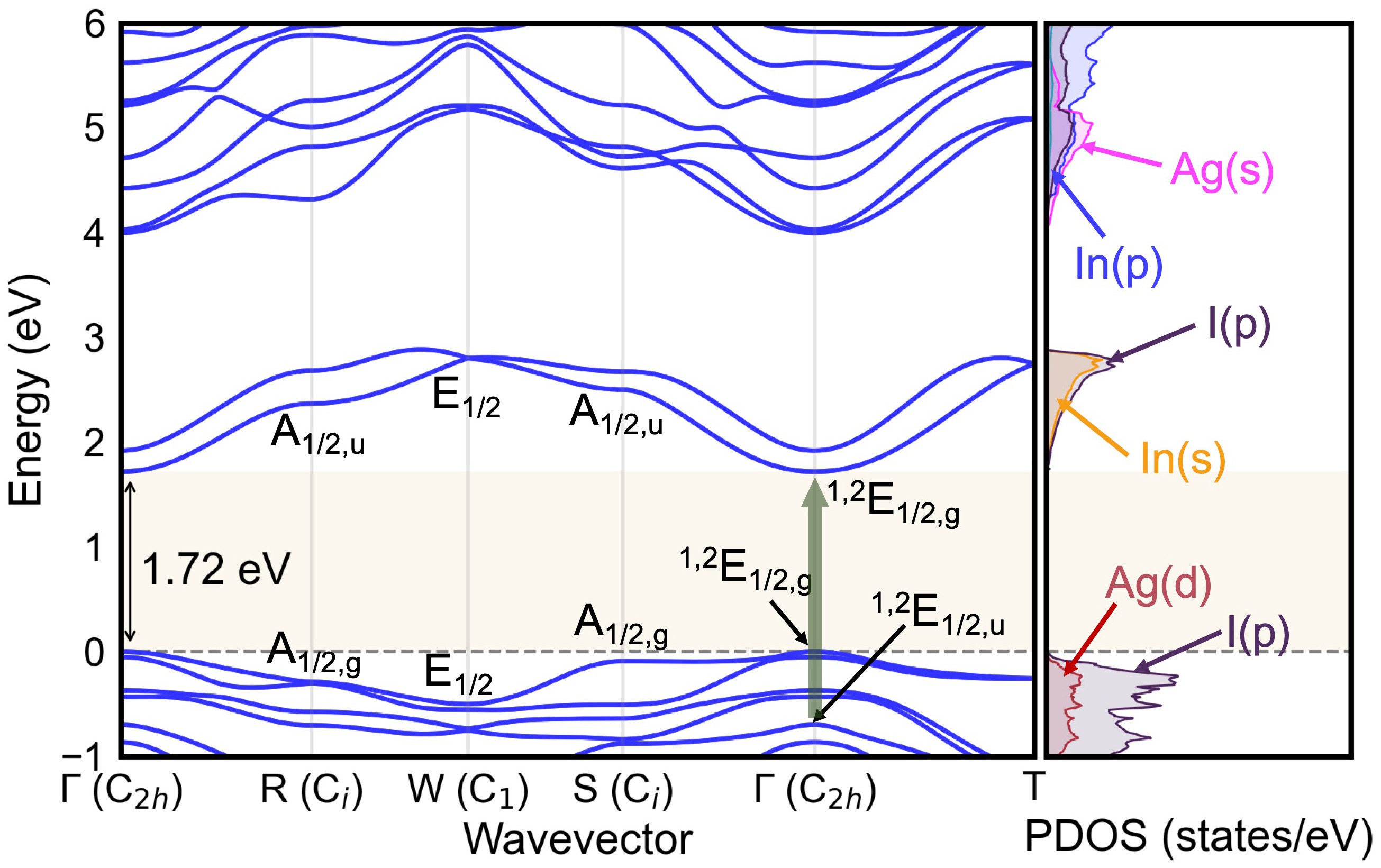}
    \caption{Electronic band structure of AgInI$_4$ based on the recuded symmetry model and the corresponding density of states, green arrow represent the first allowed transition at $\Gamma$. (All details about symmetry analysis for In double salt can be found in the Supporting Information)}
    \label{fig:bands74}
\end{figure}

Our calculations reveal a direct band gap at the $\Gamma$ point. To address the well-known underestimation of band gaps within standard DFT, we performed additional calculations using the PBE0 hybrid functional (see Computational Details in the Supporting Information), obtaining a direct band gap of 1.72~eV. This behavior is similar to the indirect-to-direct transition observed in the Cs$_2$AgInCl$_6$ double perovskite relative to Cs$_2$AgBiCl$_6$, now reproduced in the AgBiI$_4$–AgInI$_4$ system. The valence band maximum (VBM) is primarily composed of I-5\emph{p} orbitals hybridized with Ag-4\emph{d} states, and its dispersion closely resembles that of the AgBiI$_4$ ternary compound. This orbital configuration and the resemblance between the two materials is expected, as the absence of Bi-\emph{s} contributions at the VBM eliminates the Ag-\emph{d}/Bi-\emph{s} interaction, as shown in Figure~\ref{fig:bands74} and Figure~S2 for AgInI$_4$ and AgBiI$_4$, respectively. This is a notable difference from halide double perovskites, as the indirect band gap in AgBiI$_4$ originates from the position of the conduction band minimum (CBM) at the R point, and not at the VBM. In AgInI$_4$, the CBM exhibits a narrower bandwidth and is formed predominantly by hybridized In-5\emph{s} and I-5\emph{p} orbitals, whereas in the Bi-based compound it arises mainly from Bi-5\emph{p} and I-5\emph{p} states (Figure~\ref{fig:bands74}). Notably, the CBM orbital character in AgInI$_4$ closely parallels that of Cs$_2$AgInCl$_6$, both featuring In-5\emph{s} and I-5\emph{p} contributions. To quantify the band dispersion, we computed the effective masses at the band edges for the In-based compound, obtaining average values of 0.75~$m_e$ for holes and 0.63~$m_e$ for electrons, where $m_e$ denotes the free electron mass. For comparison, the corresponding values for the Bi-based compound are 0.77~$m_e$ and 0.42~$m_e$ for holes and electrons, respectively~\cite{Cucco2022}. The effective masses in both In- and Bi-based ternary compounds are slightly larger than those of their double perovskite analogues~\cite{Volonakis2016,Volonakis2017}.

\subsection{Optical Properties and Photovoltaic Performance}\

\begin{figure}[H]
    \centering
    \includegraphics[width=0.99\textwidth]{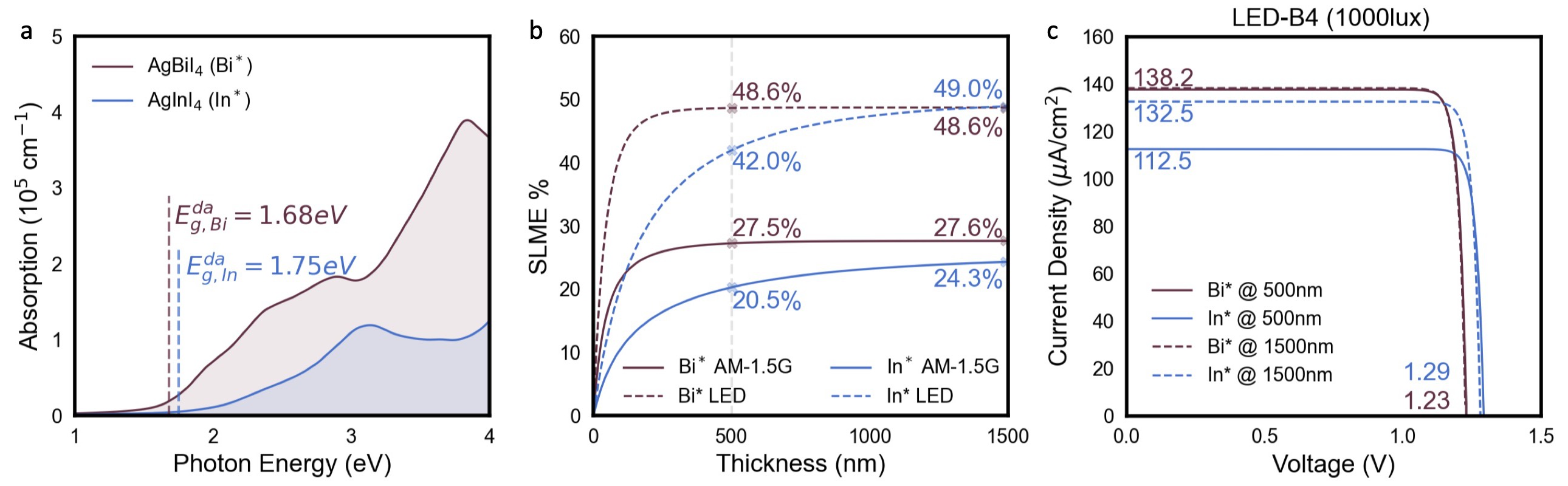}
    \caption{a) The theoretical absorption coefficient and corresponding first allowed dipole transitions for AgInI$_4$ (light blue) and AgBiI$_4$ (red). b) The SLME of double halide materials under standard solar and LED illumination. c) The corresponding J-V curves under indoor conditions.}
    \label{fig:optical}
\end{figure}

Next, we analyze the optical properties of the AgInI$_4$ halide double salt and compare them with those of AgBiI$_4$. Figure~\ref{fig:optical}a shows the calculated absorption coefficients for both compounds. The bismuth-based compound exhibits a sharper absorption onset near its calculated band gap, $E_g = 1.67$~eV, as the first allowed direct transition occurs at nearly the same energy ($E_g^{da} = 1.68$~eV)~\cite{Cucco2022}. In contrast, the indium-based salt displays a smoother, less intense absorption onset, with the first allowed transition appearing at $E_g^{da} = 1.75$~eV (Figure~S1), also close to its fundamental band gap of 1.72~eV. Notably, the first dipole-allowed transition at the $\Gamma$ point (green arrow in Figure~\ref{fig:bands74}) occurs at a significantly higher energy of 2.42~eV. Two main factors account for the weaker absorption of AgInI$_4$ compared to AgBiI$_4$. First, the transition at the direct band gap is symmetry-forbidden, resulting in a reduced absorption onset. Second, the markedly stronger absorption observed in AgBiI$_4$ above 3.0~eV can be attributed to the presence of the 6s$^2$ lone-pair electrons of Bi$^{3+}$ in the valence band (Figure~S2). This is analogous to the 6s$^2$ configuration of Pb$^{2+}$, which is known to play a key role in the distinctive optoelectronic behavior of the ABX$_3$ class of materials~\cite{fabini2020}. In contrast, AgInI$_4$ contains In$^{3+}$ cations with a filled 4d$^{10}$ shell and no active lone-pair electrons, leading to a weaker optical response.

Having established the electronic and optical properties of the In-based double salts, we now examine their potential as solar absorbers and indoor energy-harvesting materials. To assess their photovoltaic performance, we employ the spectroscopic limited maximum efficiency (SLME), which accounts for essential parameters such as the fundamental band gap, the first allowed dipole transition, and the absorption coefficient~\cite{Zunger2012} to estimate an upper limit in the potential performance of a material under different sources of illumination. As shown in Figure~\ref{fig:optical}b, AgBiI$_4$ is predicted to achieve a theoretical power conversion efficiency exceeding 27\%, in good agreement with previous computational reports~\cite{Cucco2022,Pecunia2020}. In comparison, AgInI$_4$ attains a maximum SLME of approximately 20.5\% at a film thickness of 500~nm. Increasing the film thickness to 1.5~$\mu$m reveals that the Bi-based compound reaches its performance saturation around 500~nm under solar illumination, whereas the efficiency of the In-based compound continues to increase. The superior SLME of AgBiI$_4$ , under both solar and LED illumination, can be attributed for low thicknesses to the steeper rise in its absorption coefficient. The slightly higher efficiency of AgBiI$_4$ beyond 1500~nm under solar conditions reflects its narrower band gap, which enables the absorption of near-infrared photons that are not captured by AgInI$_4$.  For completeness, we compare the SLME of AgInI$_4$ under outdoor illumination with that of other In-based semiconductors reported in the literature. The AgInI$_4$ double salt exhibits a lower SLME compared to (Ag/Cu)–In–(Te/Se/S)$_2$ chalcopyrites and CuAu-like phase compounds reported by Lamoen \emph{et al.}~\cite{Lamoen2016}, but remains comparable ($\approx$20\%) to AgInS$_2$ in the chalcopyrite phase and CuInSe$_2$ in the CuAu-like phase ($\approx$21\%). Detailed numerical data are provided in Table~S1 of the Supporting Information.

Under LED illumination, the AgBiI$_4$ double salt is predicted to achieve an efficiency of 48.6\% at a film thickness of 500~nm, consistent with the predictions by Pecunia \textit{et al.}~\cite{Pecunia2020}. In comparison, AgInI$_4$ exhibits a slightly lower efficiency of 42.0\% at the same thickness, which gradually increases with film thickness. The reduced performance of the In-based compound at small thicknesses originates from its lower absorption coefficient, consistent with the trend observed under solar illumination. Nevertheless, under LED illumination, both materials display comparable performance at larger thicknesses, as the photon energy of the LED source lies well above their optical band gaps (see Figure~S4). When benchmarked against other direct band gap materials reported in the literature, the reduced-symmetry AgInI$_4$ model demonstrates performance comparable to MAPbI$_3$ (49.1\%), InI (53.4\%), and Sb$_2$S$_3$ (51.1\%), all of which possess similar optical band gaps. Moreover, AgInI$_4$ outperforms materials with band gaps deviating more substantially from the optimal range for indoor photovoltaic applications (1.8–2.0~eV), such as CsSnI$_6$ (36.5\%), (MA)$_3$Sb$_2$I$_9$ (39.6\%), and (FA)$_3$Bi$_2$I$_9$ (33.4\%) (see Table~S2). Figure~\ref{fig:optical}c shows the current density–voltage (J–V) curves under LED illumination at 500~nm. The Bi-based compound achieves a current density of 138.2~$\mu$A\,cm$^{-2}$, slightly exceeding that of the In-based compound (112~$\mu$A\,cm$^{-2}$). While the current density of AgBiI$_4$ remains nearly unchanged with increasing absorber thickness, AgInI$_4$ exhibits a 15\% increase. The predicted open-circuit voltages (V$_\mathrm{oc}$) under LED illumination are 1.23~V and 1.29~V for the Bi- and In-based PIMs, respectively. These values are nearly identical, reflecting the comparable band gaps of the two double halide salts. A similar trend is observed under solar illumination (Figure~S5, Supporting Information).

Although the AgInI$_4$ structure is predicted to be stable both dynamically and against decomposition into its constituent salts, no experimental reports of this compound exist to date. Furthermore, in our preliminary ambient pressure solution attempts (see Supplementary Notes 1 and 2 in SI for the details), films prepared from 1:1 AgI:InI$_3$ precursors were poorly crystalline (two weak reflections) and displayed nearly featureless visible absorption (see SI, Figure S6). Hot injection at 185 $^{\circ}$C produced light-yellow particles whose powder XRD did not match the simulated AgInI$_4$ pattern (SI, Figure S7). Taken together with our symmetry analysis where the $\Gamma$ bright transition lies well above the fundamental gap ($\Gamma$ bright $\approx$ 2.42 eV vs Eg $\approx$ 1.72 eV) and the small computed formation enthalpy of the WPS model (-12 meV/atom), these observations indicate that phase pure, crystalline AgInI$_4$ is difficult to isolate via the solution routes explored here. In contrast, analogous AgBiI compounds readily crystallized under the same synthetic conditions~\cite{Ghosh2018,Matuhina2023}, indicating that further optimization of precursor chemistry and processing parameters may be necessary.  our initial synthesis attempts did not yield crystalline AgInI$_4$. Consequently, we proceed with an in-depth computational exploration of the Ag–In–I phase space. To this end, we perform a systematic high-throughput materials screening using the Materials Project database~\cite{Jain2013}, as described in the following section.

\subsection{Ag/In phase space screening}\

We first established our screening criteria by considering materials with the same stoichiometric ratio (1:1:4) as AgInI$_4$, while excluding those containing actinide or lanthanide elements. This initial search yielded 1,636 available structures. We further filtered the dataset to include only materials in which the A- and B-site atoms exhibit octahedral, tetrahedral, or mixed tetrahedral–octahedral (spinel-type) coordination environments. This choice reflects the coordination characteristics of the predicted AgInI$_4$ structure, as well as those of the binary precursors AgI and InI$_3$. Next, we restricted the selection to materials that are both experimentally reported and thermodynamically stable, resulting in a subset of 78 potential crystal structures. Owing to the smaller atomic radius and higher electronegativity of oxygen compared to halogens and other chalcogens—which generally lead to more compact and strongly bonded lattices—we further excluded oxides from the dataset. This refinement produced a final set of 23 compounds (Table~S4), which were used as structural templates to explore potentially stable competing phases (i.e., polymorphs) within the unexplored Ag–In–I phase space. The chemical stability of all candidate polymorphs was assessed by calculating their formation enthalpies, $\Delta H$, with respect to the binary salts AgI and InI$_3$ (Figures~S8a and S8b), under ambient conditions. For the substitutional engineering study we set as the reference structure the initial Wyckoff position splitting crystal, which we call polymorph-1 (p1) thereafter. 

\begin{figure}[H]
    \centering
    \includegraphics[width=0.99\linewidth]{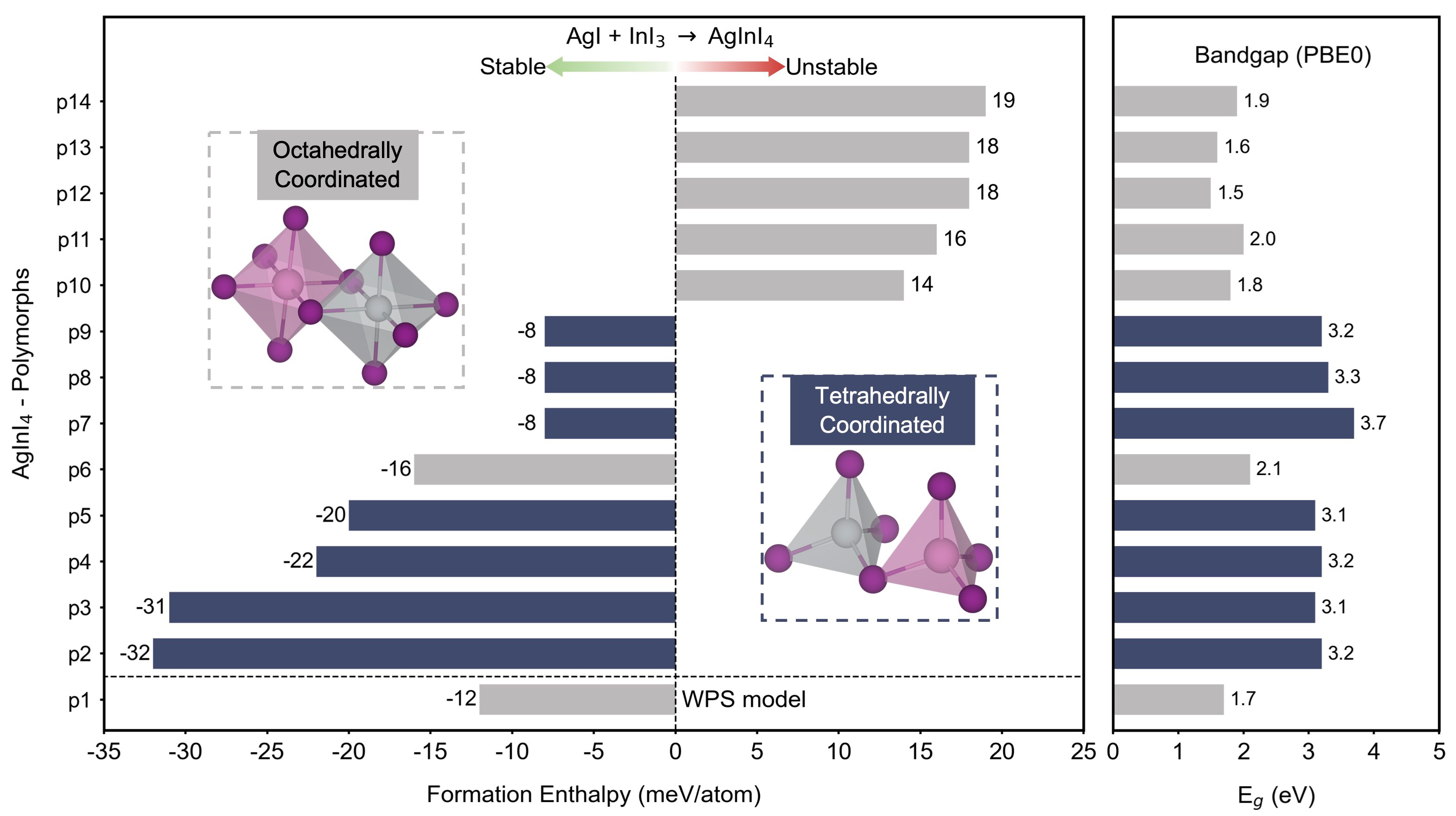}
    \caption{Formation energy of AgInI$_4$ polymorphs in respect to AgI and In$_3$ salts and the correspoding coordination environment of each polymorph (left). Band gap (E$_g$) of each polymorph employing PBE0 functional (right). p1 corresponds to polymorph obtained by Wyckoff position splitting.}
    \label{fig:MatScreening}
\end{figure}

Figure~\ref{fig:MatScreening} presents the formation energies of the most stable polymorphs obtained from (Ag:In) / (In:Ag) substitutions within the original materials set. Additional details on the screening process and structural analysis are provided in the *Materials Screening* section of the Supporting Information (Figures~S8–S12). Among the 13 additional unique polymorphs identified within the Ag–In–I phase space, we find five (p10 to p14) to be unstable, with formation energies ranging from +14 to +19 meV/atom. In addition, we identify several stable polymorphs, some of them, such as p7 to p9, are less energetically favorable than the reference p1, with formation enthalpies close -8 meV/atom. Most interestigly, we find a set of polymorphs that exhibit better chemical stability than the reference structure, with formation energies ranging from -16 to -32 meV/atom. This indicates that p1, given its dynamic stability, might indeed be a metastable phase within the Ag-In-I chemical space. Since our exploration begins from structural templates featuring octahedral, tetrahedral, and mixed (spinel-type) coordination environments, we next analyzed the local coordination around the A and B sites, while also performing Ag–In and In–Ag substitutions in the screened crystal structures. Interestingly, even when the initial configurations were derived from spinel-type structures, the resulting geometries consistently relaxed into either fully tetrahedrally or fully octahedrally coordinated frameworks, as illustrated in Figure~S12. This outcome allows us to classify the identified compounds into two distinct families: tetrahedrally coordinated (blue) and octahedrally coordinated (gray) structures, as summarized in Figure~\ref{fig:MatScreening}. Overall, we observe a clear trend in which tetrahedrally coordinated phases exhibit greater thermodynamic stability than their octahedral counterparts. This can be rationalized by the fact that both parent salts, AgI and InI$_3$, adopt tetrahedral coordination environments under ambient conditions. We further note that this trend is likely pressure-dependent: under high-pressure conditions, both AgI~\cite{Hull1999} and InI$_3$~\cite{Ni2024} transition to octahedrally coordinated phases (Figure~S8c and S8d).

\begin{figure}[H]
    \centering
    \includegraphics[width=0.99\linewidth]{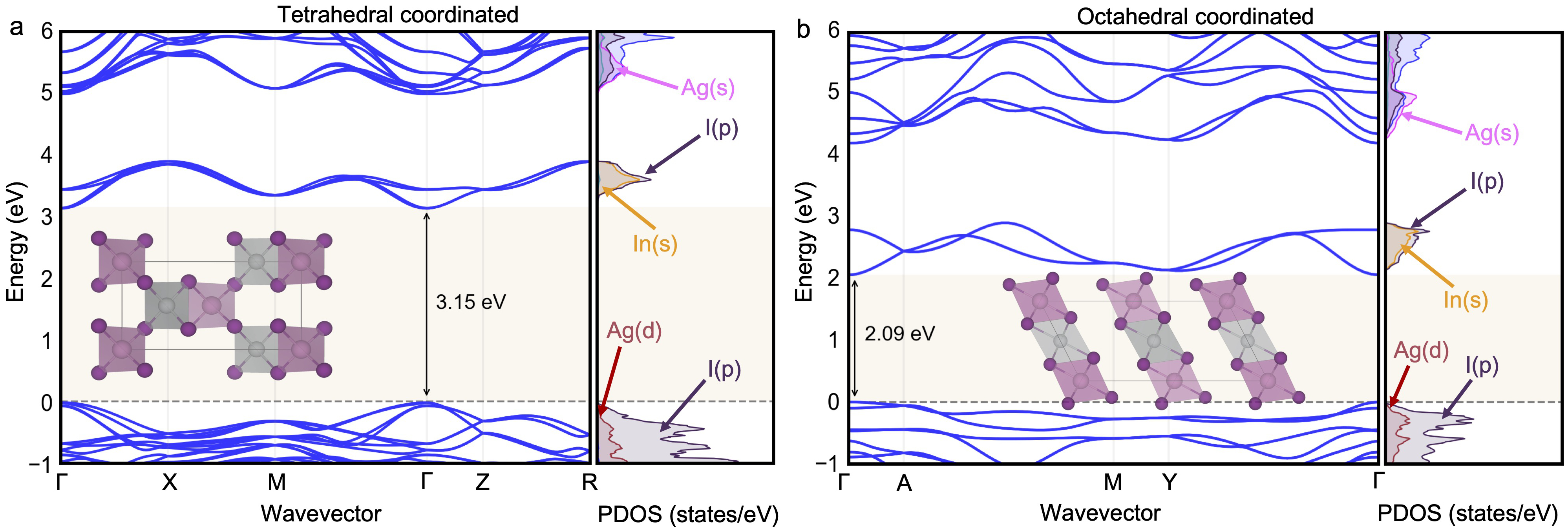}
    \caption{Electronic structure of the most stable polymorphs of each family. a) Presents the electronic structure of the tetrahedral coordinated polymorph (Polymorph-2), while b) panel displays the electronic structure of the octahedral coordinated polymorph (Polymorph-6).(Crystal structure insets depic the conventional cell of each polymorph).}
    \label{fig:Polybands}
\end{figure}

Finally, we evaluate the electronic structure of the identified polymorphs, with their calculated band gaps summarized in Figure~\ref{fig:MatScreening}. Interestingly, all polymorphs featuring tetrahedral coordination at both cation sites exhibit wide band gaps of approximately 3.0~eV. In contrast, the octahedrally coordinated phases display significantly smaller gaps, typically around 2.0~eV, as obtained using the PBE0 hybrid functional, which is known to provide reliable estimates for halide perovskites~\cite{Garba2025}. For the two most stable representatives of each coordination type, the tetrahedral polymorph (p2) and the octahedral polymorph (p6), we calculated the full electronic band structures shown in Figure~\ref{fig:Polybands}. The corresponding band gaps are 3.15~eV for p2 and 2.09~eV for p6. In both cases, the VBM is primarily composed of I-5$p$ orbitals with notable hybridization from Ag-4$d$ states, consistent with the electronic structure of p1 (Figure~\ref{fig:bands74}). The valence band of the tetrahedral p2 polymorph exhibits a markedly higher dispersion compared to the flatter bands near the $\Gamma$ point observed for the octahedral phases p6 and p1. To quantify this trend, we computed the average hole effective masses, obtaining values of 0.52~$m_e$ for p2, 0.75~$m_e$ for p1, and 1.13~$m_e$ for p6. Similarly, the average electron effective masses at the CBM, which are dominated by hybridized In-5$s$ and I-5$p$ orbitals, are 0.32~$m_e$, 0.34~$m_e$, and 0.63~$m_e$ for p2, p1, and p6, respectively (Table~S3, Supporting Information). Among the octahedrally coordinated structures, the most stable polymorph adopts a layered edge-sharing octahedral configuration (inset of Figure~\ref{fig:Polybands}b), exhibiting a direct band gap of 2.09~eV. By comparison, p1, characterized by a three-dimensional network of edge-sharing octahedra (Figure~\ref{fig:Structure}), shows a slightly smaller gap of 1.72~eV. Overall, the wide band gaps of the tetrahedrally coordinated polymorphs, particularly p2, render them unsuitable as photovoltaic absorbers but potentially attractive as electron transport layers or more generally as wide-band-gap semiconductors due to their relatively low effective masses. On the other hand, the octahedrally coordinated phases exhibit direct band gaps in the visible range and may therefore serve as promising candidates for light-emitting applications. This simple structure–property link provides a useful guide for targeting band gaps through coordination control (Coordination–gap rule). However, within halide double salts, substituting pnictogens with indium appears detrimental to visible-light absorption and thus to photovoltaic performance, as compared with their Bi or even Sb-based analogues.

\section{Conclusion}
In summary, we have presented a comprehensive computational investigation of the Ag–In–I phase space, beginning with substitutional engineering of the reduced-symmetry model of AgBiI$_4$ through the introduction of In$^{3+}$. The resulting compound, AgInI$_4$, is predicted to be both chemically and dynamically stable, exhibiting a direct band gap analogous to that of the well-known Cs$_2$AgInCl$_6$ double perovskite. However, the spectroscopic limited maximum efficiency (SLME) of AgInI$_4$ is found to be lower than that of other materials with comparable band gaps reported in the literature. We attribute this reduction primarily to symmetry-forbidden transitions and, more importantly, to the absence of the Bi-derived lone-pair 6s$^2$ electrons at the valence band maximum, which typically enhance s–p optical transitions. To expand the search for stable phases, we performed a high-throughput screening of compounds with the same 1:1:4 stoichiometry using the Materials Project database~\cite{Jain2013}. This screening led to the identification of several new stable and metastable Ag–In–I polymorphs, which can be categorized into two structural families based on their local coordination environments. A detailed analysis of their electronic properties revealed a clear correlation between coordination geometry and band gap: octahedrally coordinated polymorphs exhibit band gaps around 2.0~eV, while tetrahedrally coordinated ones display wider band gaps near 3.0~eV. It is also worth noting that the compound predicted to be the most stable within the Ag–In–I phase space (p2) is also identified as a stable crystal structure for this chemical composition by graph neural network predictions~\cite{merchant2023}. Finally, we emphasize that the Wyckoff position splitting (WPS) method represents a powerful approach for constructing atomistic models from experimental structures. Nonetheless, it should be applied with caution as a predictive tool in substitutional engineering studies, as it does not exhaustively sample the full configurational phase space of unexplored materials. 

Overall, our exploration of the Ag–In–I chemical space narrows the prospects of these materials as efficient photovoltaic absorbers, yet highlights their potential as non-toxic alternatives for charge transport layers or for other optoelectronic applications requiring materials with tunable band gaps from visible-range emitters to wide-band-gap or transparent semiconductors. Tetrahedral polymorphs ($\approx$ 3 eV) appear more suited to transport/window layers, whereas octahedral phases ($\approx$ 2 eV) may be explored for emitters or indoor PV admixtures subject to synthesis. Future work should test symmetry-breaking or alloying routes to brighten the $\Gamma$-point transitions and mechanochemical/high-pressure syntheses to access metastable octahedral Ag–In–I polymorphs. The insights gained here provide a foundation for further experimental and theoretical work aimed at broadening the design landscape of halide-based semiconductors beyond the conventional lead-based and pnictogen chemistry.
 

\medskip
\textbf{Acknowledgements}\par
 C.T., M.K. and G.V. acknowledge the Agence Nationale pour la Recherche through the CPJ program and the SURFIN project (ANR-23-CE09-0001), and the ALSATIAN project (ANR-23-CE50-0030), and the ANR under the France 2030 programme, MINOTAURE project (ANR-22-PETA-0015). P.V. and G.K.G. acknowledge that the work is part of the Research Council of Finland Flagship Programme, Photonics Research and Innovation (PREIN), Decision No. 346511. Access to the HPC resources of TGCC was obtained under the Allocation Grant No. 2025 - A0190907682 made by GENCI. We acknowledge computational resources from the EuroHPC Joint Undertaking and supercomputer LUMI [https://lumi-supercomputer.eu/], hosted by CSC (Finland) and the LUMI consortium through a EuroHPC Extreme Scale Access call. 

\medskip
\textbf{Author contributions}\par
All authors participated in the preparation of the manuscript.


\medskip
\textbf{Competing financial interests}\par
The authors declare that they have no competing financial interests.

\medskip
{\bf References\\}
\bibliographystyle{naturemag}
\bibliography{ref.bib}

\end{document}